\documentclass[pra,twocolumn,a4paper,showpacs]{revtex4}
\usepackage{amsmath}
\usepackage{amssymb}
\usepackage{amsfonts}
\usepackage{graphicx}
\usepackage{hyperref}
\usepackage{bm}

\begin{document}
\title{Wave-packet analysis of interference patterns in output coupled atoms}

\author{Kari~H\"{a}rk\"{o}nen}
\email{Kari.Harkonen@utu.fi}
\altaffiliation{Current address: Department of Physics and Astronomy, University of Sussex, Brighton BN1 9QH, UK} 
\affiliation{Turku Centre for Quantum Physics, Department of Physics and Astronomy, University of Turku, FI-20014 Turku, Finland} 

\author{Otto~Vainio}
\affiliation{Turku Centre for Quantum Physics, Department of Physics and Astronomy, University of Turku, FI-20014 Turku, Finland} 

\author{Kalle-Antti~Suominen}
\affiliation{Turku Centre for Quantum Physics, Department of Physics and Astronomy, University of Turku, FI-20014 Turku, Finland} 

\date{\today}

\begin{abstract}
We study the output coupling of atoms from a magnetic trap into a linear potential slope of gravity using a weak radio-frequency field. We present a one-dimensional wave-packet model based on a continuous loading of a continuous spectrum of generalised eigenstates to describe the scenario. Analyzing the model, we show how the interference of the classical coupling fields maps to the interference of the resulting atomic streams.
\end{abstract} 


\maketitle


\section{Introduction}

Ever since the first realization of an atomic Bose--Einstein condensation
\cite{Anderson1995,Davis1995,Bradley1995,Pethick2002,Pitaevskii2003},
there have been applications, where the coherent cloud of trapped
atoms has been used as a source for output coupling \cite{Mewes1997}.
The coherence properties of the source can be mapped to a coherent
output \cite{Andrews1997,Oettl2005} and, moreover, by applying continuous
coupling a coherent stream of spatially wide-spread atoms can be created
\cite{Bloch1999,Koehl2001}. In a close analogy with the optical laser,
an atom laser is thus formed.

An often used method to realise such a system, is to induce spin flips
to the trapped cloud of atoms by introducing a weak magnetic field
perturbation, i.e., an oscillating radio frequency (rf) field perpendicular
to the static trapping field \cite{Mewes1997,Band1999,Gerbier2001}.
The rf-field creates a coupling between the Zeeman sublevels $M_{F}$
and the internal spin state can be thus flipped to an untrapped or,
with strong rf-field intensities, even to anti-trapped states \cite{Vitanov1997,Robins2005}.
Especially, in the linear Zeeman shift regime, the sublevel $M_{F}=0$
does not couple to the static trapping magnetic field at all, but
is only affected by the linear potential slope of the gravity. Consequently,
such atoms fall freely and exit the trapping area. Other implementations
of the output coupling are, e.g., to apply a Raman transition \cite{Hagley1999},
which could also provide the free-falling atomic flux with an initial
momentum kick, or to construct a tunneling connection \cite{Anderson1998}.
Interestingly, the output coupling situation is reminiscent of the
molecular dissociation triggered by ultrashort pulses \cite{Suominen1992,Garraway1995,Paloviita1997}. 

Over the years, there have been theoretical papers considering the
atom lasers with one-dimensional \cite{Schneider1999,Schneider2000,Dugue2007}
and three-dimensional models \cite{Gerbier2001,Kramer2002,Kramer2006},
using weak \cite{Band1999,Japha1999,Choi2000} and strong \cite{Band1999,Haine2002,Haine2003,Dugue2007}
coupling strengths, applying multiple simultaneous couplings \cite{Schneider2000,Kramer2006},
having the source at finite temperature \cite{Japha1999,Choi2000},
and from the point of view of stability \cite{Robins2001,Haine2002,Haine2003}
and pulse shape \cite{delCampo2007}. In this paper we analyze a simple
one-dimensional model in order to clarify one specific problem concering
interference patterns due to multiple simultaneous couplings.

The phase coherence of the spatially elongated atomic beams is most
strikingly demonstrated by strong interference patterns while superimposing
two beams with different energies \cite{Bloch2000,Bourdel2006,Vainio2006}.
Again in a close analogy with the optical lasers, the interference
pattern depends on three quantities: (i) the relative amplitudes,
(ii) the relative phase difference, and (iii) the energy separation.

The spatially wide-spread wave-function interpretation of the interfering
atomic beams is a strongly non-classical result. However, an alternative
explanation in terms of interfering (classical) magnetic rf-fields,
which are driving the coupling, has been proposed \cite{Vainio2006}.
In this line of reasoning the coupling magnetic field is understood
in terms of a carrier frequency and a beating envelope, and the correspondence
between the pulsing rf-amplitude and the resulting output stream was
demonstrated. There seems to be a discrepancy between the two ways
of looking at the problem. On the one hand, the system is described
by a pulsing flux generated by a pulsing semi-classical coupling,
while on the other hand, the system is described by interference of
superimposed spatially elongated asymptotic atomic wavefunctions \cite{Bloch2000,Schneider2000}.
The purpose of this paper is to demonstrate the connection between
these two extreme interpretations.

In Sec.~\ref{sec:derivation} we derive a wave-packet solution to
a simplified one-dimensional problem in terms of a continuous loading
of a continuous spectrum of generalised energy eigenstates. In Sec.~\ref{sec:visibility}
we show how the visibility of the atomic interference pattern maps
from the interference of the magnetic fields. We then apply the model
in Sec.~\ref{sec:application} using realistic experimental parameters
and compare the results with numerical simulations including the complete
Zeeman-sublevel structure as well as the atomic contact interactions.
Finally, we finish with conclusions and discussion in Sec.~\ref{sec:conclusions}.


\section{\label{sec:derivation}Wave packet model}


\subsection{Physical system}

An atom couples to the magnetic field via its magnetic moment, resulting
in an interaction energy defined as 
\begin{equation}
U(\mathbf{B})=-\boldsymbol{\mu}\cdot\mathbf{B}.
\label{eq:magneticMoment}
\end{equation}
Above, the magnetic moment operator is $\boldsymbol{\mu}=-\mu_{0}(g_{S}\mathbf{S}+g_{L}\mathbf{L}+g_{I}\mathbf{I})/\hbar$,
where $\mu_{0}=|e|\hbar/2m_{e}$ is the Bohr magneton and $g_{i}$
are the Landé $g$-factors for electronic spin ($S$), orbital ($L$),
and nuclear spin ($I$) angular momentum. When the energy splitting
corresponding ~to this term is small compared to fine and hyperfine
splittings, the total angular momentum $\mathbf{F}=\mathbf{I}+\mathbf{J}$,
with $\mathbf{J}=\mathbf{L}+\mathbf{S}$, is a good quantum number
and $\boldsymbol{\mu}\simeq-\mu_{0}g_{F}\mathbf{F}/\hbar$, where
the Landé factor is $g_{F}\simeq g_{J}[F(F+1)+J(J+1)-I(I+1)]/2F(F+1)$,
with $g_{J}\simeq1+[J(J+1)+S(S+1)-L(L+1)]/2J(J+1)$. In the limit
of weak magnetic field, which is the case in the present work, the
Zeeman splitting between the sublevels $M_{F}$ is linear \cite{Bransden2003}.


\subsubsection{Trapping potential and gravity}

A magnetic trap for the atoms in the low-field-seeking states is formed
by simply creating a magnetic field intensity minimum. The local direction
of the field describes the quantization axis $\hat{\mathbf{e}}_{z}$,
and close to the minimum, the magnetic field is assumed to be approximately
harmonic, such that $\mathbf{B}_{\textrm{trap}}=B_{\textrm{trap}}(\mathbf{r})\hat{\mathbf{e}}_{z}=B_{\textrm{trap}}^{0}(\lambda_{x}^{2}x^{2}+\lambda_{y}^{2}y^{2}+\lambda_{z}^{2}z^{2})\hat{\mathbf{e}}_{z}$.
In the same direction, a strong static bias field $\mathbf{B}_{\textrm{bias}}=B_{\textrm{bias}}\hat{\mathbf{e}}_{z}$
is applied in order to remove the degeneracy at origin and, hence,
to suppress the Majorana spin flips and the resulting atom losses
\cite{Pitaevskii2003}.

The trapping potential operator is $U_{\textrm{trap}}(\mathbf{r})=\textrm{sgn}(g_{F})[\frac{1}{2}m(\omega_{x}^{2}x^{2}+\omega_{y}^{2}y^{2}+\omega_{z}^{2}z^{2})+\hbar\omega_{\textrm{bias}}]F_{z}/\hbar$,
where $\omega_{i}^{2}=2\mu_{0}|g_{F}|B_{\textrm{trap}}^{0}\lambda_{i}^{2}/m$
and $\omega_{\textrm{bias}}=\mu_{0}|g_{F}|B_{\textrm{bias}}$. The
atoms feel, irrespective of their internal state, also the linear
potential of the gravity, $U_{\textrm{gravity}}(\mathbf{r})=-mgx$;
the harmonic trapping potentials are relocated accordingly in position
and energy. The static Hamiltonian reads 
\begin{equation}
H_{0}=T+U_{\textrm{trap}}(\mathbf{r})+U_{\textrm{gravity}}(\mathbf{r}),
\label{eq:H0}
\end{equation}
where $T=-\hbar^{2}\nabla^{2}/2m$ is the kinetic energy term. As
is now obvious, an integer-valued hyperfine state $F$ supports a
special sublevel $M_{F}=0$, which feels only the linear gravitational
potential.


\subsubsection{Coupling rf-field}

The coupling between the Zeeman sublevels is induced by applying a
weak rf-field 
\begin{equation}
\mathbf{B}_{\textrm{rf}}(t)=\frac{1}{2}B_{0}(t)\hat{\mathbf{e}}_{\mathrm{rf}}e^{-i(\omega_{\textrm{rf}}t+\theta)}+c.c.
\end{equation}
with a finite component in the direction perpendicular to the trapping
field. The pulse envelope $B_{0}(t)$ has an arbitrary shape and the
pulse is turned on after the initial time $t=0$. As will be clear
from the following, the model can be generalised directly to any linear
combination of such single-mode rf-fields. Consequently, it is sufficient
now to consider a single rf-field.

The rf-field results in an interaction Hamiltonian 
\begin{equation}
H_{I}(t)=-\boldsymbol{\mu}\cdot\mathbf{B}_{\textrm{rf}}(t).
\label{eq:Hint}
\end{equation}
We will write the polarization vector as $\hat{\mathbf{e}}_{\textrm{rf}}=\sum_{i=+,-,z}(\hat{\mathbf{e}}_{i}\cdot\hat{\mathbf{e}}_{\textrm{rf}})\hat{\mathbf{e}}_{i}$,
where $\hat{\mathbf{e}}_{\pm}=(\hat{\mathbf{e}}_{x}\pm i\hat{\mathbf{e}}_{y})/\sqrt{2}$.
The $z$-component causes only a small perturbation to the trapping
potential, and is assumed to be zero hereafter. The circular components,
corresponding to the raising and lowering angular momentum operators
$F_{\pm}=F_{x}\pm iF_{y}$, induce transitions between the sublevels,
as $F_{\pm}|F,M_{F}\rangle=\hbar\sqrt{F(F+1)-M_{F}(M_{F}\pm1)}|F,M_{F}\pm1\rangle$.
Finally, we remark that the total angular momentum operator $F^{2}$
commutes with the total Hamiltonian $H(t)=H_{0}+H_{I}(t)$ as well
as its components $H_{0}$ and $H_{I}(t)$, so the dynamics is confined
into a single hyperfine state $F$.


\subsection{Representation of the state}

The coupling is assumed to be weak, so only transitions to sublevels
$M_{F,\mathrm{final}}=M_{F,\textrm{initial}}\pm1$
are relevant. Since the magnetic (trapping) potentials for the different
sublevels are $\langle F,M_{F}|U_{\textrm{trap}}|F,M_{F}\rangle\propto\textrm{sgn}(g_{F})M_{F}$,
we will assume that the trapped atoms are initially on the internal
state $|T\rangle\equiv|F,M_{F}=\textrm{sgn}(g_{F})\rangle$ and that
the free-falling untrapped state is $|U\rangle\equiv|F,M_{F}=0\rangle$.
The transition between the internal states $|T\rangle\to|U\rangle$
is provided by the operator $F_{+/-}$ in systems with negative/positive
$g_{F}$.

\begin{figure}[tb!]
\begin{centering}
\includegraphics{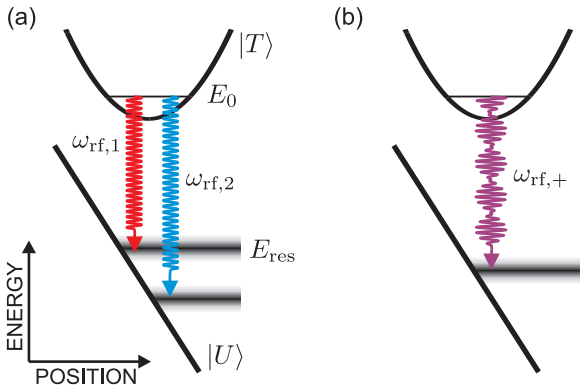} 
\end{centering}
\caption{\label{fig:physicalSystem} (Color online) Schematic setup of potentials
and couplings. The trapped Zeeman sublevel $|T\rangle$ is coupled
to the untrapped level $|U\rangle$ by a weak rf-field. Starting from
a trapped state with energy $E_{0},$ the rf-frequency determines
the resonance energy $E_{\textrm{res}}$, around which a continuous
spectrum of generalised energy eigenstates is populated during the
output coupling. (a) With multiple simultaneous frequency components
$\omega_{\textrm{rf},j}$, a corresponding set of resonant energy
levels is formed. The atomic streams interfere as they fall in the
gravity field. (b) Equivalently, the coupling can be interpreted as
being driven by the sum of the single rf-fields, which correspond
to a carrier frequency $\omega_{\textrm{rf},+}=(\omega_{\textrm{rf},1}+\omega_{\textrm{rf},2})/2$
and a coupling strength pulsing at frequency $\omega_{\textrm{rf},-}=(\omega_{\textrm{rf},1}-\omega_{\textrm{rf},2})/2$.}
\end{figure}

In the following, we will neglect any population transfer to the antitrapped
high-field-seeking Zeeman sublevels [$M_{F}=-n\,\textrm{sgn}(g_{F})$,
with $n=1,\ldots,F$], for which the magnetic field minimum forms
a repulsive potential, as well as to the more energetic trapped sublevels
[$M_{F}=n\,\textrm{sgn}(g_{F})$, with $n=2,\ldots,F$]; the relevant
potentials and couplings are illustrated in Fig.~\ref{fig:physicalSystem}.
This is a justified neglection since we are interested in the weak
coupling regime. On the other hand, if a maximal flux of atoms would
be desirable, one would have to use strong rf-fields, and in that
case these neglected sublevels would have a nontrivial contribution
\cite{Robins2005}. With ever stronger coupling strengths the system
should be described by dressed potentials \cite{Zobay2001,Zobay2004,Schumm2005,Hoffenberth2006}.

The convenient choice of basis functions depends on the internal state.
For the trapped state $|T\rangle$, the basis is provided by the harmonic
oscillator eigenstates $\{|\phi_{n}\rangle\}_{n=0}^{\infty}$. For
the untrapped state $|U\rangle$, however, the external potential
is linear, and the basis is formed by an uncountable set of generalised
eigenfunctions, or distributions, $\{|\psi_{E}\rangle\}_{E\in\mathbb{R}}$,
which satisfy the Airy differential equation \cite{Abramowitz1970}.
Their explicit form is 
\begin{equation}
\psi_{E}(x)=\mathcal{N}\textrm{Ai}\big{[}(x+E/mg)/l\big{]},
\end{equation}
where the normalisation factor $\mathcal{N}=1/l\sqrt{mg}$ and the
characteristic length scale $l=(\hbar^{2}/2gm^{2})^{1/3}$. These
generalised functions are not normalisable according to the $L^{2}$
norm, $\langle\psi_{E}|\psi_{E'}\rangle=\delta(E-E')$, and thus can
not individually represent any physical state. However, they form
a complete orthonormal spatial basis, in a sense that $\int\textrm{d}E\,\psi_{E}^{*}(x)\psi_{E}(x')=\delta(x-x')$.
Therefore, any spatial state $|\varphi\rangle$ can be described in
terms of these distributions as $|\varphi\rangle=\int\mathrm{d}E\, f(E)|\psi_{E}\rangle$,
where the spectrum is $f(E)=\langle\psi_{E}|\varphi\rangle$. Consequently,
the normalisation of the state is done in accordance with the properties
of the spectrum, such that $\|\varphi\|^{2}=\int\mathrm{d}E\,|f(E)|^{2}$.
It is immediately evident, that any state $|\varphi_{D},U\rangle$
described by a discrete spectrum $f(E)=\sum_{i}c_{i}\delta(E-E_{i})$
is, first of all, unphysical, and corresponds to a (quasi)periodic
solution within the evolution generated by $H_{0}$. This does not
fit to the intuition about a free-fall event. Combining the previous
statements, any state within our system can be expressed as
\begin{equation}
|\Psi(t)\rangle=\sum_{n}b_{n}(t)|\phi_{n},T\rangle+\int\mathrm{d}E\, c_{E}(t)|\psi_{E},U\rangle.
\label{eq:stateRepresentation}
\end{equation}
In the interaction picture with respect to $H_{0}$, given by Eq.~\eqref{eq:H0},
$|\tilde{\Psi}(t)\rangle=e^{iH_{0}t/\hbar}|\Psi(t)\rangle$ and the
corresponding coefficients are $\tilde{b}_{n}(t)=e^{iE_{n}t/\hbar}b_{n}(t)$
and $\tilde{c}_{E}(t)=e^{iEt/\hbar}c_{E}(t)$. In the following, we
will also use the notation $|\Psi_{\beta}\rangle\equiv\langle\beta|\Psi\rangle$,
where $\beta=T,U$, for the trapped and untrapped components of the
total state.


\subsection{\label{sec:wavepacket} Wave-packet solution}

The coupling between the different sublevels comes from the interaction
Hamiltonian~\eqref{eq:Hint}. The weak coupling causes only a small
perturbation to the bare system, defined by the static Hamiltonian~\eqref{eq:H0},
and therefore its effect can be described by transition matrix elements.
Let us assume that initially the system is at equilibrium in a trapped
ground state $|\Psi(0)\rangle=|\phi,T\rangle$, for which $H_{0}|\phi\rangle=E_{0}|\phi\rangle$.
In terms of the representation~\eqref{eq:stateRepresentation}, the
coefficients are $b_{0}(0)=1$, $b_{n}(0)=c_{E}(0)=0$ for all $n>0$
and $E\in\mathbb{R}$.

In the interaction picture the equation of motion for the coefficients
$\tilde{c}_{E}(t)$ is given by 
\begin{align}
\frac{\mathrm{d}}{\mathrm{d}t}\tilde{c}_{E}(t)= & \langle\psi_{E},U|\frac{\mathrm{d}}{\mathrm{d}t}\tilde{\Psi}(t)\rangle=-\frac{i}{\hbar}\langle\psi_{E},U|\tilde{H}_{I}(t)|\tilde{\Psi}(t)\rangle\nonumber \\
= & -\frac{i}{\hbar}\tilde{b}_{0}(t)e^{-i(E_{0}-E)t/\hbar}\langle\psi_{E}|\phi\rangle\langle U|H_{I}(t)|T\rangle.
\label{eq:componentDerivative}
\end{align}
The trapped state remains essentially intact during the weak coupling
pulse, so we can assume $\tilde{b}_{0}(t)=1$. The formal solution
in the Schr\"{o}dinger picture is 
\begin{align}
c_{E}(t)= & -\frac{i}{\hbar}e^{-iEt/\hbar}\langle\psi_{E}|\phi\rangle\int_{0}^{t}\textrm{d}s\, e^{-i(E_{0}-E)s/\hbar}\nonumber \\
 & \times\langle U|H_{I}(s)|T\rangle.
\end{align}
Therefore, according to definition~\eqref{eq:stateRepresentation},
the untrapped component is given by 
\begin{align}
|\Psi_{U}(t)\rangle= & -\frac{i}{\hbar}\int_{0}^{t}\textrm{d}s\, e^{-iE_{0}s/\hbar}\langle U|H_{I}(s)|T\rangle\nonumber \\
 & \times\int\textrm{d}E\, e^{-iE(t-s)/\hbar}\langle\psi_{E}|\phi\rangle|\psi_{E}\rangle.
\end{align}
Defining the outcoupling rate function $\Omega$ and the respective
instantaneous outcoupled state $|\Phi\rangle$, corresponding to a
delta-peak outcoupling rate function, as 
\begin{align}
\Omega(t) & \equiv-\frac{i}{\hbar}e^{-iE_{0}t/\hbar}\langle U|H_{I}(t)|T\rangle, 
\label{eq:outcouplingRateFunction}\\
|\Phi(t)\rangle & \equiv\int\textrm{d}E\, e^{-iEt/\hbar}\langle\psi_{E}|\phi\rangle|\psi_{E}\rangle=\langle U|e^{-iH_{0}t/\hbar}|\phi,U\rangle, 
\label{eq:instantaneousOutcoupledState}
\end{align}
the full time-dependent solution for the outcoupled atomic beam can
be written in a compact form as a convolution 
\begin{equation}
|\Psi_{U}(t)\rangle=\int_{0}^{t}\textrm{d}s\,\Omega(s)|\Phi(t-s)\rangle=\big{[}\Omega\ast(\Theta|\Phi\rangle)\big{]}(t),
\label{eq:convolution}
\end{equation}
where the Heaviside theta function, for which $\Theta(t)$ equals
to 0 for $t<0$ and 1 for $t>0$, takes care of a proper temporal
causality. The instantaneous outcoupled state $|\Phi(t)\rangle$ matches
the static-Hamiltonian-induced evolution [cf. Eq.~\eqref{eq:H0}]
of the spatial component of the initial trapped state $|\phi\rangle$,
only its internal state is the untrapped one. Finally, we remind that
$\Omega(t)$ vanishes for $t<0$ according to our previous definition.


\subsection{Continuous spectrum of states}

Let us consider the matrix element of the interaction Hamiltonian
between the trapped and untrapped states $\langle U|H_{I}|T\rangle$.
Since the trapped and the untrapped states are separated by a single
quantum of angular momentum, $\langle T|F_{z}|T\rangle=\textrm{sgn}(g_{F})\hbar=\pm\hbar$
and $\langle U|F_{z}|U\rangle=0$, the transition between the states
$|T\rangle\to|U\rangle$ is induced by the operator $F_{\alpha}$,
where $\alpha=+/-$ for systems with negative/positive $g_{F}$. Therefore,
Eq.~\eqref{eq:componentDerivative} is 
\begin{align}
\frac{\mathrm{d}}{\mathrm{d}t}\tilde{c}_{E}(t)= & -\frac{i}{2\sqrt{2}\hbar^{2}}\mu_{0}g_{F}B_{0}(t)\langle\psi_{E}|\phi\rangle\langle U|F_{\alpha}|T\rangle\nonumber \\
 & \times\big{[}(\hat{\mathbf{e}}_{\alpha}\cdot\hat{\mathbf{e}}_{\textrm{rf}})e^{-i[(E_{0}+\omega_{\textrm{rf}}-E)t+\theta]}\nonumber \\
 & \quad\,\,\,+(\hat{\mathbf{e}}_{\alpha}\cdot\hat{\mathbf{e}}_{\textrm{rf}}^{*})e^{-i[(E_{0}-\omega_{\textrm{rf}}-E)t-\theta]}\big{]},
\end{align}
where the factor $\sqrt{2}$ comes from the identity $\hat{\mathbf{e}}_{\pm}\cdot\mathbf{F}=F_{\pm}/\sqrt{2}$. 

With a constant rf-field, $B_{0}(t)=B_{0}\Theta(t)$, the time integration
gives terms 
\begin{equation}
\tilde{c}_{E}(t)\propto t\langle\psi_{E}|\phi\rangle\textrm{sinc}\Big{(}\frac{E_{0}\pm\omega_{\textrm{rf}}-E}{2}t\Big{)}.
\end{equation}
Therefore, the spectrum concentrates in the vicinity of resonant energy
levels $E=E_{0}\pm\omega_{\textrm{rf}}$ as time passes. According
to the physical setup, on the other hand, the overlap integral $\langle\psi_{E}|\phi\rangle$
is concentrated around $E\simeq-mg\langle\phi|x|\phi\rangle\ll E_{0}$.
Consequently, the significant contribution accumulates around the
resonant energy level 
\begin{equation}
E_{\textrm{res}}\equiv E_{0}-\omega_{\textrm{rf}}.
\label{eq:Eres}
\end{equation}
In terms of the generalised eigenstates, there will always be a continuous
range of occupied states around the resonant energy $E_{\textrm{res}}$.


\section{\label{sec:visibility}Visibility of the interference pattern}

The form of the free-falling atomic cloud $|\Psi_{U}\rangle$ was
in Eq.~\eqref{eq:convolution} expressed as a convolution of the
outcoupling rate function $\Omega(t)$ and a spatial term $|\Phi(t)\rangle$.
Next we will consider the emerging interference patterns due to multiple
rf-fields driving the coupling simultaneously.

The corresponding (classical) magnetic field components $\mathbf{B}_{\textrm{rf}}^{i}(t)$
interfere with each other, such that the total field is $\mathbf{B}_{\textrm{rf}}(t)=\sum_{i}\mathbf{B}_{\textrm{rf}}^{i}(t)$.
Correspondingly, the outcoupling rate function $\Omega(t)=\sum_{i}\Omega_{i}(t)$
and, because of the linearity of Eq.~\eqref{eq:convolution}, the
outcoupled component is 
\begin{align}
|\Psi_{U}(t)\rangle & =[\Omega\ast(\Theta|\Phi\rangle)](t)=\sum_{i}[\Omega_{i}\ast(\Theta|\Phi\rangle)](t)\nonumber \\
 & =\sum_{i}|\Psi_{U}^{i}(t)\rangle.
\label{eq:interference}
\end{align}
The interference pattern appears similarly in the (quantum) matter
fields as a sum of atomic streams, each of which corresponds to an
atomic beam outcoupled by a single rf-field component.


\subsection{\label{sec:classicalInterference}Interference of classical fields}

The point of view expressed in Ref.~\cite{Vainio2006} was that the
combination of the (classical) magnetic fields, which operate at frequencies
$\omega_{1}$ and $\omega_{2}$ with equal constant amplitudes, corresponds
to a single field, whose carrier frequency is the average $\omega_{+}=(\omega_{1}+\omega_{2})/2$
and the pulse envelope is modulated at frequency $\omega_{-}=(\omega_{1}-\omega_{2})/2$
(cf. Fig.~\ref{fig:physicalSystem}). Moreover, the relative phase
difference between the circular components driving the outcoupling,
\begin{equation}
\Delta\theta=\textrm{arg}(\hat{\mathbf{e}}_{\alpha}\cdot\hat{\mathbf{e}}_{\textrm{rf,1}}^{*}e^{i\theta_{1}})-\textrm{arg}(\hat{\mathbf{e}}_{\alpha}\cdot\hat{\mathbf{e}}_{\textrm{rf,2}}^{*}e^{i\theta_{2}}),
\label{eq:relativePhaseDifference}
\end{equation}
shifts the envelope of the interference pattern and, consequently,
the intensity profile of the falling stream of atoms. Generally, the
interference pattern depends on (i) the relative amplitudes, (ii)
the relative phase difference, and (iii) frequency separation. 

Based on this description, one might expect that whenever the carrier
frequency $\omega_{+}$ falls into the region where the overlap integral
$|\langle\psi_{E_{0}-\hbar\omega_{+}}|\phi\rangle|$ is finite, there
would be a finite stream of atoms falling from the trap, and the intensity
of the stream would be modulated at frequency $\omega_{-}$, such
that the maxima of the rf-field coincide with the maxima of the atomic
intensity (see Fig.~3 in Ref.~\cite{Vainio2006}).


\subsection{\label{sec:quantumInterference}Interference of quantum fields}

The wave-packet result derived in Sec.~\ref{sec:wavepacket} explains
why the above-mentioned simplistic analogy from the classical interference
is not exactly true. According to Eq.~\eqref{eq:interference}, the
visibility of the interference pattern is affected by two contributions:
(i) interference pattern of the magnetic fields and (ii) convolution
by the temporal free-fall evolution of the initial trapped state profile.
Looking at the stream at a particular position $x$ as a function
of time, the interference pattern of the magnetic fields, possibly
with perfect visibility, is smoothed by the temporal width of the
instantaneous outcoupled state $|\Phi(x,t)\rangle$ falling past this
point. 

For a Gaussian initial state $|\phi_{0}\rangle$, the analytical solution
\begin{equation}
|\Phi(x,t)\rangle\propto\exp\Big{[}-\frac{(x-x_{0}-\frac{1}{2}gt^{2})^{2}}{2\sigma(t)^{2}}\Big{]},
\end{equation}
where $x_{0}=g/\omega^{2}$ and $\sigma(t)=\sqrt{\sigma_{0}^{2}+t^{2}/\sigma_{0}^{2}}$,
with $\sigma_{0}=\sqrt{\hbar/m\omega}$, allows us to estimate the
temporal width. Namely, at time $t$ the wave packet has a spatial
width of $\sigma(t)$ centralised around position $x_{0}+\frac{1}{2}gt^{2}$,
and the center of mass falls with velocity $v(t)=gt$, so the passing
time is approximately $\sigma(t)/v(t)\ge1/\sigma g$; this value corresponds
to a balance between dispersion and gravitational acceleration. Therefore,
even with an infinitely long coupling time, the interference pattern
is still smoothed by a distribution with a finite width.

With a single rf-field, the amplitude of the falling atomic flux depends
on the applied rf-frequency $\omega_{\textrm{rf}}$. This can be seen
from the resonance energy condition~\eqref{eq:Eres} as compared
to the overlap integral $\langle\psi_{E}|\phi\rangle$. Therefore,
two different rf-frequencies generally produce atomic streams with
different amplitudes, even if the rf-field amplitudes are the same.
According to Eq.~\eqref{eq:interference}, the relative phase difference
of two magnetic fields~\eqref{eq:relativePhaseDifference} maps directly
to the relative phase difference of the resulting matter waves. This
was explicitly demonstrated in the experiment of Ref.~\cite{Vainio2006}.


\section{\label{sec:application}Application and comparison to numerical results}

In the following, we will concentrate on $^{87}$Rb atoms and in particular
its hyperfine ground state $F=1$. In this case the Landé factor is
$g_{F}=-1/2$ and, therefore, the trapped low-field-seeking Zeeman
sublevel is $|T\rangle=|F=1,M_{F}=-1\rangle$ and the untrapped one
is $|U\rangle=|F=1,M_{F}=0\rangle$ (see Fig.~\ref{fig:physicalSystem}).
The physical parameters are adopted from Ref.~\cite{Vainio2006}
and are summarised in Table~\ref{tab:parameters}.

\begin{table}[tb!]
\caption{\label{tab:parameters}The physical parameters used in the examples.
Here, $a_{0}=5.5\times10^{-11}$~m is the Bohr radius.}
\begin{ruledtabular}
\begin{tabular}{lcc} 
Quantity  & Symbol  & Value \\ \hline  
Trap frequency ($x$ and $z$ direction)  & $\omega_{x,z}/2\pi$  & 160 Hz\\
Trap frequency ($y$ direction)  & $\omega_{y}/2\pi$  & 6.7 Hz\\ 
Rabi frequency & $|\Omega|/2\pi$  & 50 Hz\\ 
Bias frequency  & $\omega_{\textrm{bias}}/2\pi$  & 900 kHz \\ 
Number of atoms  & $N$  & 10$^{5}$ \\ 
Scattering length & $a$  & 103 $a_{0}$ \\ 
\end{tabular}
\end{ruledtabular}
\end{table}

In this section we will compare the wave-packet solution~\eqref{eq:convolution}
to numerical simulations including all the Zeeman sublevels. Especially,
we show the impact of the atomic contact interactions by solving the
corresponding Gross-Pitaevskii equation \cite{Pethick2002,Pitaevskii2003}.
Both the model and the simulations are one dimensional. The contact
interactions appear as an additional non-linear mean-field term $U_{\textrm{int}}(x,t)=g_{\mathrm{1D}}|\Psi(x,t)|^{2}$.
The scaled interaction coefficient is $g_{\mathrm{1D}}=(\sqrt{\omega_{1}\omega_{2}}m/2\pi\hbar)g_{3D}$
\cite{Petrov2004}, where the three-dimensional interaction term is
$g_{\mathrm{3D}}=4\pi\hbar^{2}aN/m$, with scattering length $a$
and number of particles $N$ \cite{Pethick2002,Pitaevskii2003}.

\begin{figure}[tb!]
\begin{centering}
\includegraphics{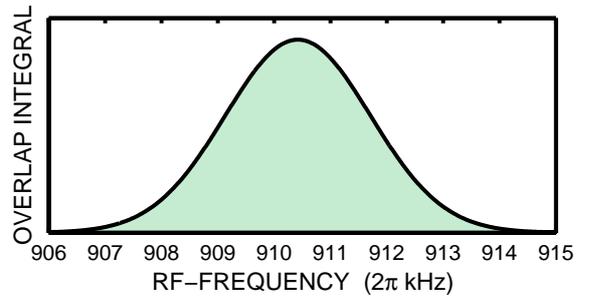} 
\end{centering}
\caption{\label{fig:overlap} (Color online) The overlap integral $\langle\psi_{E_{0}-\hbar\omega_{\textrm{rf}}}|\phi_{0}\rangle$,
in arbitrary units, as a function of rf-frequency $\omega_{\textrm{rf}}$.
As mentioned in the text, the form is well approximated by a Gaussian
shape centralised at $(E-mgx_{0})/h\simeq910.3$~kHz with width $mg\sigma/h\simeq1.8$~kHz.}
\end{figure}

When neglecting the atomic contact interactions, the ground state
of the harmonic trapping potential is a Gaussian $|\phi_{0}\rangle$.
The overlap integral between the Gaussian state and the generalised
energy eigenstates $|\psi_{E}\rangle$ can be calculated analytically
\cite{Paloviita1997,Kramer2006}. In the limit of a steep gravity
slope, $g\gg0$, the generalised energy eigenfunction approaches Dirac's
delta distribution as $|\psi_{E}(x)\rangle\sim\delta(x+E/mg)/\sqrt{mg}$.
Since the width of the trapped state $\sigma$ clearly exceeds the
characteristic length scale of the Airy distribution $l$, the overlap
integral is well approximated by $\langle\psi_{E}|\phi_{0}\rangle\simeq[\pi(mg\sigma)^{2}]^{-1/4}\exp[-(E+mgx_{0})^{2}/2(mg\sigma)^{2}],$
as is obvious in Fig.~\ref{fig:overlap}.

\begin{figure}[tb!]
\begin{centering}
\includegraphics{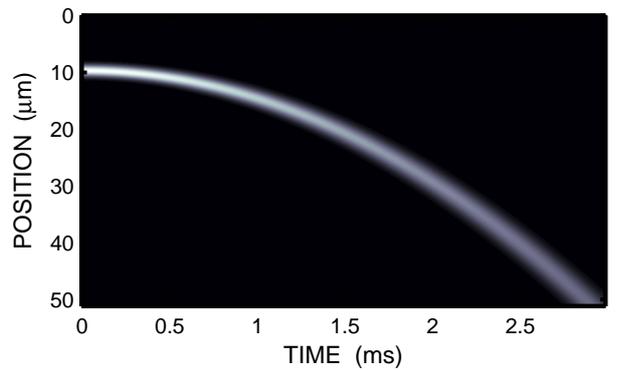}
\end{centering}
\caption{\label{fig:loadingStateEvolution} (Color online) The density profile
of the instantaneous outcoupled state $|\Phi(x,t)|^{2}$ (arbitrary
units). Initially the state is the Gaussian ground state of an harmonic
potential. In the linear potential slope the functional form is maintained,
while the center of mass accelerates according to classical mechanics,
$x_{0}(t)=x_{0}+\frac{1}{2}gt^{2}$, and the width disperses, $\sigma^{2}(t)=\sigma_{0}^{2}+t^{2}/\sigma_{0}^{2}$. }
\end{figure}

\begin{figure}[tb!]
\begin{centering}
\includegraphics{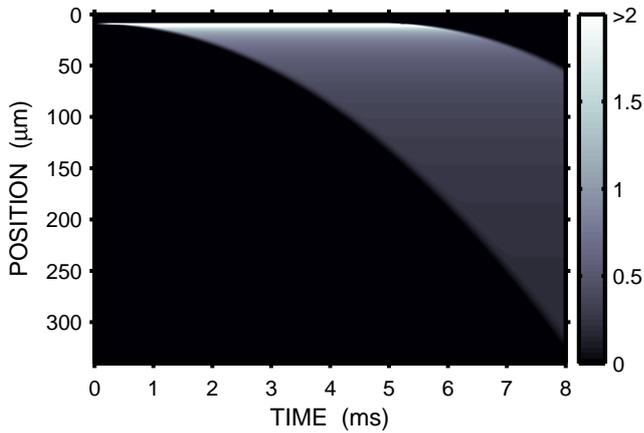}
\end{centering}
\caption{\label{fig:1rf_timeEvolution} (Color online) Output-coupled atomic
density for a 5-ms-long box-shaped pulse with rf-frequency $\omega_{\textrm{rf}}/2\pi=910$~kHz.
The density is in units of $10^{3}$~1/m and is flattened from above
to increase the clarity.}
\end{figure}

\begin{figure}[tb!]
\begin{centering}
\includegraphics{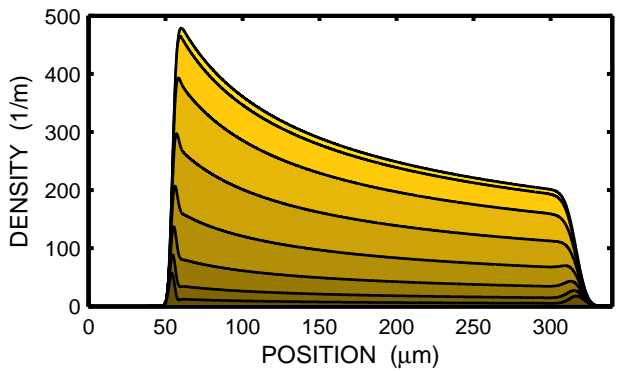}
\end{centering}
\caption{\label{fig:1rf} (Color online) Output-coupled atomic density profiles
8 ms after the beginning of a 5-ms-long box-shaped pulse. The plot
shows 8 different rf-frequencies; from lowest to highest density:
$\omega_{\textrm{rf}}/2\pi=907,907.5,\ldots,910.5$~kHz (cf. Fig.~\ref{fig:overlap}).}
\end{figure}

In Fig.~\ref{fig:loadingStateEvolution} the time evolution of the
instantaneous outcoupled state~\eqref{eq:instantaneousOutcoupledState}
for the non-interacting case is shown. The state corresponds to falling
atoms outcoupled by an infinitesimally short rf-pulse. The total wave
packet due to an rf-pulse with a finite duration, is achieved by integrating
this state over time, in accordance with Eq.~\eqref{eq:convolution}.
In the examples, we use a 5-ms-long box-shaped pulse form for the
rf-field. The field amplitude is such that the maximum Rabi frequency
is $|\Omega|/2\pi=50$~Hz. However, in the spirit of Ref.~\cite{Vainio2006},
we assume linear polarisation and the coupling is therefore suppressed
by a factor $1/\sqrt{2}$. The density profile of the resulting stream
of outcoupled atoms vs. time is shown in Fig.~\ref{fig:1rf_timeEvolution}.
The number of outcoupled atoms, as well as the density profile depends
on the applied rf-frequency. This dependency is shown in Fig.~\ref{fig:1rf}.
The increase in the density follows the amplitude of the overlap integral
shown in Fig.~\ref{fig:overlap}.

\begin{figure}[tb!]
\begin{centering}
\includegraphics{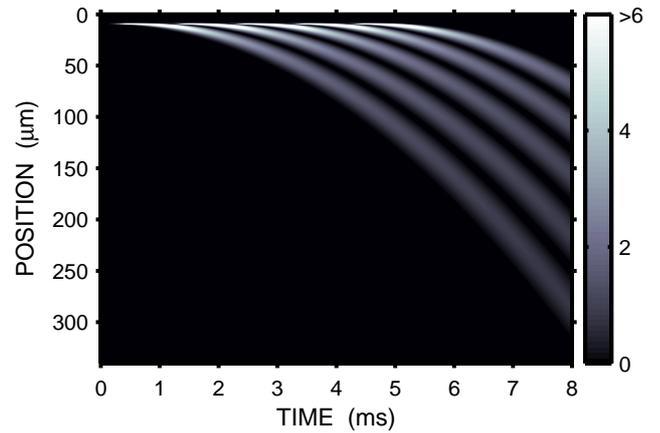}
\end{centering}
\caption{\label{fig:2rf_timeEvolution} (Color online) As Fig.~\ref{fig:1rf_timeEvolution}
but with two simultaneous equally strong pulses with rf-frequencies
$\omega_{\textrm{rf,1}}/2\pi=910$~kHz and $\omega_{\textrm{rf,1}}/2\pi=911$~kHz,
and with a relative phase difference of $\pi$.}
\end{figure}

\begin{figure}[tb!]
\begin{centering}
\includegraphics{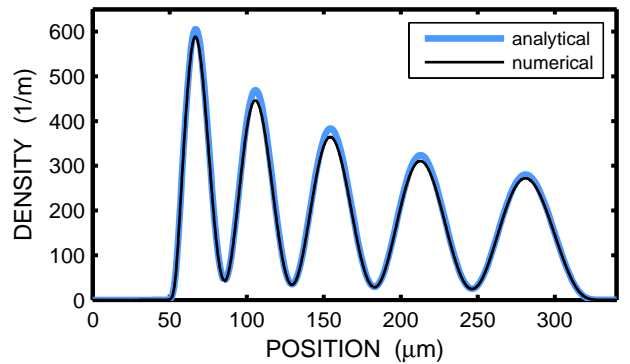}
\end{centering}
\caption{\label{fig:2rf_pi} (Color online) As Fig.~\ref{fig:1rf}, but with
two simultaneous rf-fields with $\omega_{\textrm{rf},1}/2\pi=909$~kHz
and $\omega_{\textrm{rf},2}/2\pi=908$~kHz, and a relative phase
difference of $\pi$. Analytical model (thick line) agrees well with
numerical simulation (thin line).}
\end{figure}

The density profile of a pulsating outcoupled stream, which is produced
by two simultaneous resonant rf-pulses separated in frequency by $\Delta\omega_{\textrm{rf}}/2\pi=1$~kHz
and in phase by $\Delta\theta=\pi$, is shown in Fig.~\ref{fig:2rf_timeEvolution}
as a function of time. In Fig.~\ref{fig:2rf_pi} we compare the analytically
calculated density profile to the numerically computed one. In the
numerical computation the time-dependent Schr\"{o}dinger equation
was evolved taking into account three states, i.e., one harmonically
trapped, one harmonically anti-trapped, and one affected by a linear
potential with the slope corresponding to the gravity. The numerical
computations were done for both interacting and non-interacting cases.
Overall, we find good agreement between the analytical and the numerical
results. 

\begin{figure}[tb!]
\begin{centering}
\includegraphics{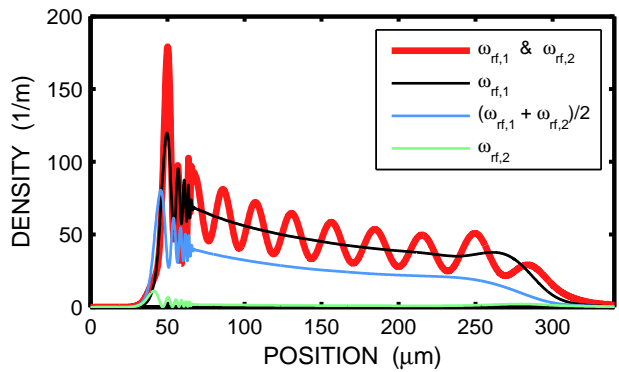}
\end{centering}
\caption{\label{fig:2rf_pi_interacting}(Color online) As Fig.~\ref{fig:1rf},
but including the atomic contact interactions. Looking at the outcoupling
near the edge of the distribution, there is a situation, where one
rf-frequency ($\omega_{\textrm{rf},1}/2\pi=903$~kHz, highest thin
straight line) produces a strong stream of atoms, another rf-frequency
($\omega_{\textrm{rf},2}/2\pi=901$~kHz, lowest thin straight line)
with the same field amplitude almost nothing, and the average rf-frequency
$(\omega_{\textrm{rf},1}+\omega_{\textrm{rf},2})/2$ (middle thin
straight line) again a clear stream. Applying both equally strong
fields simultaneously, with a relative phase difference of $\pi$,
shows interference with a limited visibility (thick oscillating line)
in accordance with the non-interacting examples. }
\end{figure}

Due to the atomic contact interactions the trapped ground state is
broadened from a Gaussian into a Thomas-Fermi distribution. Accordingly,
the range of rf-frequecies capable of producing outcoupling changes.
In Fig.~\ref{fig:2rf_pi_interacting} we show how also in the interacting
case the visibility of the interference pattern due to two equally
strong rf-fields is not perfect. Especially, our example shows the
interesting case, where one of the rf-frequencies outcouples hardly
any stream while the other one and the average do. As in the non-interacting
case, applying the equally strong fields simultaneously produces interference
with low visibility. 

\begin{figure}[tb!]
\begin{centering}
\includegraphics{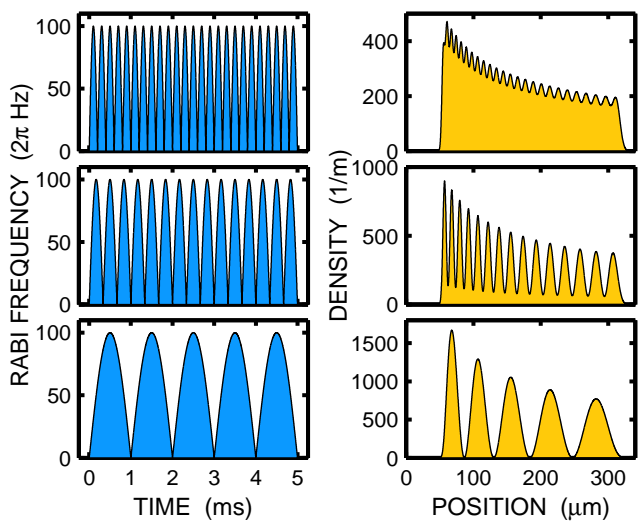}
\end{centering}
\caption{\label{fig:2rf_rf_vs_psi} (Color online) The perfect interference
patterns of the coupling magnetic fields (left panels) map to smoothed
interference in the corresponding atomic density (right panels) because
of the finite spatial extend of the trapped state. The used frequencies
are $\omega_{\textrm{rf},1}/2\pi=911$~kHz and (i) $\omega_{\textrm{rf},2}/2\pi=906$~kHz
(first row) (ii) $\omega_{\textrm{rf},2}/2\pi=908$~kHz (second row),
and (iii) $\omega_{\textrm{rf},2}/2\pi=910$~kHz (third row). The
relative phase difference between the 5-ms-long pulses is $\pi$ and
the atomic density is plotted at time $t=8$~ms.}
\end{figure}

Finally, in Fig.~\ref{fig:2rf_rf_vs_psi}, we compare the rf-pulses
and induced outcoupling streams. The figure shows clearly how the
visibility in the outcoupled atomic stream diminishes with increasing
frequency separation in the causative outcoupling rf-fields, even
if the rf-field itself is with perfect visibility. 

\section{\label{sec:conclusions}Conclusions and discussion}

We have derived a linear wave-packet solution for output coupling
scenario. The model establishes a bridge between two different ways
of looking at the interference of overlapping atom lasers, and shows
that the effect can be understood equally as interference of spatially
extended atomic clouds as well as interference of classical magnetic
fields causing the output coupling.

The model is built in terms of generalised energy eigenstates of a
linear potential caused by gravity, and it shows how the total wave
packet can be interpreted as being constructed by a continuous loading
of a continuous spectrum of these states, which individually do not
correspond to a physical solution. In general, our model does not
suffer from unphysical infinite quantities \cite{Bloch2000,Schneider2000}.

Through the analysis of the solution, it was shown that the visibility
of the observed interference pattern is limited by the spatial extend
of the trapped cloud, which serves as a source for the atomic beams.
Furthermore, the visibility is shown to be affected by the rf-frequencies
in the sense of selecting a resonant energy and, moreover, amplitude
for the atomic stream.

The simple linear model was then compared to numerical simulations
including the atomic interactions as well as all the Zeeman sublevels,
and the qualitative match was shown to be excellent using experimentally
realistic parameters. The model is one dimensional and assumes weak
coupling. The applicability is therefore restricted to cases, where
the transversal extent of the source condensate is wide \cite{Band1999,Kramer2002,Kramer2006}.
Within these restrictions, the presented linear model also generalises
straightforwardly for multiple dimensions and to an outcoupling scenario
based on a Raman transition including an initial momentum kick.


\begin{acknowledgments}
Financial support from the National Graduate School of Modern Optics
and Photonics (K.H.), the Turku University Foundation (K.H.), the
Finnish Cultural Foundation (O.V.), and the Academy of Finland (Projects
No. 115682 and 122595) is gratefully acknowledged. Also, we wish to
thank the XMDS team (www.xmds.org) for providing means for fast and
easy numerical computing.
\end{acknowledgments}



\end{document}